\documentstyle[twoside,fleqn,espcrc2,epsfig]{article}

\newcommand{\be}{\begin{equation}}
\newcommand{\ee}{\end{equation}}
\newcommand{\bea}{\begin{eqnarray}}
\newcommand{\eea}{\end{eqnarray}}
\newcommand{\setl}{\setlength\arraycolsep{2pt}}

\newcommand{\noi}{\noindent}
\newcommand{\nn}{\nonumber}
\newcommand{\ra}{\rightarrow}
\newcommand{\Ra}{\Rightarrow}

\newcommand{\lesssim}{ {\
\lower-1.2pt\vbox{\hbox{\rlap{$<$}\lower5pt\vbox{\hbox{$\sim$}}}}\ } 
}
\newcommand{\gtrsim}{ {\
\lower-1.2pt\vbox{\hbox{\rlap{$>$}\lower5pt\vbox{\hbox{$\sim$}}}}\ } 
}

\newcommand{\cG}{{\cal G}}

\newcommand{\cL}{{\cal L}}

\newcommand{\cO}{{\cal O}}

\newcommand{\cR}{{\cal R}}
\newcommand{\cS}{{\cal S}}

\newcommand{\cW}{{\cal W}}

\newcommand{\tr}{\mbox{\rm tr}}

\newcommand{\GeV}{\mbox{\rm GeV}}

\newcommand{\als}{\alpha_{\mbox{\rm {\scriptsize s}}}}

\newcommand{\muhad}{\mu_{\mbox{\rm {\scriptsize had.}}}}

\newcommand{\gL}{\frac{1-\gamma_{5}}{2}}
\newcommand{\gR}{\frac{1+\gamma_{5}}{2}}

\newcommand{\AmS}{{\protect\the\textfont2
  A\kern-.1667em\lower.5ex\hbox{M}\kern-.125emS}}

\title{\Large \bf 
$B_K$ in the chiral limit within the $1/N_c$ expansion.
\thanks{UAB-FT-496. Talk delivered at the Int. Euroconference 
on Quantum Chromodynamics, Montpellier, France, Jul 2000.}}

\author{Santiago Peris\address{Grup de Fisica Te{\`o}rica and IFAE, 
Universitat Aut{\`o}noma de Barcelona, \\
        E-08193 Bellaterra, Barcelona, Spain}}

\begin{document}

\begin{abstract}
I report on a recent calculation done in collaboration with E. de Rafael 
\cite{BK} of the invariant $\hat{B}_{K}$ factor of
$K^{0}$--$\bar{K^{0}}$ mixing in the chiral limit and to next--to--
leading order in the $1/N_c$ expansion. This calculation is, to the best of
our knowledge, the first example of a calculation of $\hat B_K$ in which there
is an explicit analytic cancellation of the $\mu$ renormalization scale and
the scheme dependence between the Wilson coefficient and the corresponding
kaon matrix element. I try to emphasize the ideas involved in the approach
and how the method could be applied to other physical situations, rather than
the details of the numerical analysis for which I refer the reader to
ref. \cite{BK}.  

\end{abstract}

\maketitle

\section{INTRODUCTION}

\noi
Real understanding of weak interactions of kaons is always hindered by the
fact that their QCD interactions are not perturbative in $\als$. 
This is a
problem with (at least) 
two well separated scales (namely $M_W$ and $\Lambda_{QCD}\sim 1$ GeV) and,
therefore,  
the well known technique of Effective Lagrangians can be applied. 
There is no problem 
as long as one is in the perturbative regime of the
expansion in $\als$, but kaons are lighter than $\Lambda_{QCD}$ and
therefore this perturbative regime is long gone by the time one gets down to
the kaon scale. In practice one assumes that the effective Lagrangian
obtained after integrating the heavy fields of the Standard Model
(i.e. heavier than $\Lambda_{QCD}$) is valid even immediately below the charm
threshold. The problem then is to match this effective Lagrangian valid at
$\mu \lesssim m_c$ to the effective Lagragian valid at $\mu \sim M_K$, wich is
the relevant one.

To be more specific, integrating heavy fields in the Standard
Model results in an effective $\Delta S=2$ Lagrangian that reads
  
\bea\label{effhal} 
{\cL}_{\rm eff}^{\Delta S=2}\!\!\!\!\!\! &=\!\!\! \!\! 
&-\frac{G_{F}^{2}M_{W}^2}{4\pi^2}
\left[\lambda_{c}^2F_1+\lambda_{t}^2F_2+2\lambda_{c}\lambda_{t}F_3\right]\nn \\
&&C_{\Delta S=2}(\mu)Q_{\Delta S=2}(x)\, ,
\eea
where $\lambda_{q}=V_{\rm qd}^{\ast}V_{\rm qs}\,,\,{\rm q}={\rm u,c,t}\,,$
with $F_{1,2,3}$ being functions of the heavy masses  of the 
fields $t$, $Z^0$, $W^{\pm}$, $b$, and $c$ which have been integrated
out~\footnote{For a detailed discussion, see e.g. refs.~\cite{Jamin,BBL96}
and references therein.} and 
\be
\label{eq:deltas2} Q_{\Delta S=2}(x)\equiv
\big[\bar{s}_{L}(x)\gamma^{\mu}d_{L}(x)\big]^2\, ,
\ee
with $q_{L}\equiv\gL q$ is a four-quark operator. This four-quark
operator has a hidden dependence on the renormalization scale 
because its matrix elements are divergent. This
divergence is subtracted away using the ${\overline {MS}}$ 
renormalization scheme,
but a $\mu$ scale dependence is left behind in the process. 
In fact, because the short
distance behavior of the operator (\ref{eq:deltas2}) is {\it not} the same as
the matrix elements of the full Standard Model where $W$, etc.., are virtually
propagating, one must introduce some coefficients, the so-called Wilson
coefficients $C_{\Delta S=2}(\mu)$ that correct for this effect. It is a
consistency condition then that the $\mu$ dependence should drop out of any
physical observable, since after all  
one could always do the calculation in the full Standard Model (at least in
principle), 
where no $\mu$ scale is necessary. Like with the $\mu$ parameter, 
the same ``declaration of independence'' 
also applies to any convention: naive 
dimensional regularization vs. 't Hooft-Veltman, definition of evanescent
operators, etc... However the question immediately arises, how can kaon matrix
elements of the operator (\ref{eq:deltas2})  know what you decided to do with
the commutator of the $\gamma_5$ matrix in the calculation of $C_{\Delta
  S=2}(\mu)$ ? For one thing kaon matrix elements of the operator
(\ref{eq:deltas2})  
will depend on masses and effective couplings but there can be no single 
Dirac matrix since kaons are bosons, of course! The dependence on
conventions has to be implicitly hidden in what you call couplings and
massess of mesons when you go away from d=4 dimensions in dimensional  
regularization. I shall explicitly demonstrate this point using the
large-$N_c$ expansion later on. I shall also show the explict $\mu$
cancellation between Wilson coefficients and matrix elements of mesons. 

At the scale $\mu\sim M_K$ chiral symmetry constraints the form of all the
possible operators appearing in the correspondimng effective Lagrangian in a 
chiral expansion in powers of derivatives and masses. To lowest order, which
is $\cO(p^2)$, there is a unique operator with the same symmetry properties as
$Q_{\Delta S=2}(x)$, which is~\cite{PdeR91,deR95}
\be
\label{CHI2}
-\frac{F_{0}^4}{4}\ g_{\Delta S=2}(\mu)\ 
\tr\left[
\lambda_{32}(D^{\mu}U^{\dagger})U\lambda_{32}(D_{\mu}U^{\dagger})U\right]
\,,
\ee
where $\lambda_{32}$ and $U$ are matrices in flavor space. 
$\lambda_{32}$ denotes the matrix $\lambda_{32}=\delta_{i3}
\delta_{2j}$, and $U$ is the $3\times 3$ unitary 
matrix which
collects the Goldstone fields and which under chiral rotations 
transforms like $U\ra V_{R}UV_{L}^{\dagger}$.  The parameter $g_{\Delta
S=2}(\mu)$ is a dimensionless coupling constant which depends on the 
underlying dynamics of spontaneous chiral symmetry breaking (S$\chi$SB) in
QCD and that has to be determined, as always, by a matching condition. 
Leaving aside the known functions $F_{1,2,3}$ it has become conventional
to define the so-called
invariant $\hat B_K$ parameter by means of the following matrix element
\be
\label{eq:Bpar} <\bar{K}^0|C_{\Delta S=2}(\mu)Q_{\Delta S=2}(0)|K^0>\equiv
\frac{4}{3}f_{K}^2M_{K}^2\hat B_{K}.
\ee
Because of the presence of $C_{\Delta S=2}(\mu)$, $\hat B_K$ so defined must
be $\mu$ (and convention) independent. There is also 
an analogous equation without the
wilson coefficient $C_{\Delta S=2}(\mu)$ that defines the associated 
$B_K(\mu)$ parameter which, unlike $\hat B_K$, is obviously $\mu$ (and
convention) dependent. From Eqs.  (\ref{eq:Bpar},\ref{CHI2}) it follows that
(with inclusion of the chiral corrections in the factorized contribution)  
\be\label{BKhat}
\hat{B}_{K}=\frac{3}{4}C_{\Delta S=2}(\mu)\times g_{\Delta S=2}(\mu)\,,
\ee 
in other words, $\hat{B}_{K}$ is also fixed by a matching condition. The rest
of this paper is a report on the calculation of $\hat{B}_{K}$ carried out in
ref. \cite{BK}. 

The Wilson coefficient is known from renormalization group-improved
perturbation theory and reads
{\setl
\bea
&&C_{\Delta S=2}(\mu)= \nn \\ \label{wilson} 
&&\!\!\!\!\left[1+\frac{\als(\mu)}{\pi}\frac{1}{\beta_{1}}\left(\gamma_2+
\frac{\gamma_1}{-\beta_1}\beta_{2}
\right)\right]
\left(\frac{1}{\als(\mu)}\right)^{\frac{\gamma_1}{-\beta_1}}\,
 \\ \label{wilsonN}
 & &\!\!\!\! \ra \! 
\left[1+\frac{\als(\mu)}{\pi}\left(\frac{1433}{1936}+\frac{1}{8}\kappa
\right)\right]
\left(\frac{1}{\als(\mu)}\right)^{\frac{9}{11}\frac{1}{N_c}}\,,
\eea}
where the explicit 
expressions for $\beta_1, \beta_2, \gamma_1$ and  $\gamma_2$ can be
found in ref.\cite{BK} and $\kappa$ encodes the choices for $\gamma_5$ in d
dimensions: $\kappa=0$ in naive dimensional regularization and $\kappa=-4$ in
the 't Hooft-Veltman scheme. Eq. (\ref{wilsonN}) is 
obtained after insisting on the
large-$N_c$ limit, which consists in taking $N_c\rightarrow \infty$ with the
condition $N_c \als\rightarrow constant$\cite{tH74}. In the strict large-$N_c$
limit one finds
\be \label{BKN}
\hat B_K =\frac{3}{4}\quad ,\quad (N_c\ra \infty) \ .
\ee

As to the large-$N_c$ expansion, reference \cite{Wi79}
explains why taking this limit is a useful thing to do because it 
simplifies QCD without crippling it; although the assumption 
that QCD confines in this limit must be made. 

For us it will be very important that, because the large-$N_c$
expansion {\it is} a systematic approximation to QCD (both in the perturbative
and nonperturbative regimes), it can be used as an interpolating function
between low energies and high energies in a systematic way. 

Our calculation will be based on the following inputs:
\begin{itemize}
\item
 The large $N_c$ expansion.
\item
 Chiral perturbation theory controls the low-energy end of QCD Green's
 functions. 
\item
 The Operator Product Expansion controls the high-energy end of QCD Green's
 functions starting from $Q\sim 1$GeV onwards.
\item
 The curve connecting low-energy and high-energy is smooth, i.e. there are no
 ``wiggles'' in the intermediate region. Such wiggles would 
destroy the cherished belief that Nature seems to be ``subtle but not
 malicious''.\footnote{There's no free lunch. {\it Any}
 interpolating procedure needs two good ends to bridge from, and 
the assumption that nothing wild is going on in between !!}
\end{itemize}

Let us now go on with the matching condition that $g_{\Delta
S=2}(\mu)$ must satisfy. From the fact that the covariant
derivatives contain external fields 
($D^{\mu}U^{\dagger}U=\partial^{\mu}U^{\dagger}U+iU^{\dagger}
r^{\mu}U -il^{\mu}$), one sees that Eq. (\ref{CHI2}) yields a
mass term for the external $r^{\mu}_{\bar d s}$ 
field\footnote{Flavor indices are important !}, among other terms. 
This $r^{\mu}_{\bar d s}$ is of course the same external
field that couples to the right-handed current $\bar d_R \gamma^{\mu} s_R$ 
appearing in the kinetic term for the quark in QCD, i.e.
\be
\label{kinetic}
\bar q \!\! \left\{\gamma^{\mu}(\partial_{\mu}\!\!+iG_{\mu})\!\!-\!
i\gamma^{\mu}\!\left[l_{\mu}\gL\!+\!r_{\mu}\gR\!\right]\!\right\} q\, . 
\ee    
Therefore at the level of the QCD Lagrangian a mass term for the 
$r^{\mu}_{\bar d s}$ field can only come about because of the operator 
$Q_{\Delta S=2}$ of Eq. (\ref{eq:deltas2}) in the Lagrangian
(\ref{effhal}). Equating the mass term $r^{\mu}_{\bar d s}r_{\mu}^{\bar d s}$
obtained from Eq. (\ref{CHI2}) to that obtained from Eq. (\ref{effhal})
(supplemented with $\cL_{\rm QCD}$ of course) one obtains the matching
condition for $g_{\Delta S=2}(\mu)$\cite{BK}:
{\setl
\bea\label{GDELTAS2}
 &&g_{\Delta S=2}(\mu,\epsilon)=
\quad 1 - \nn \\
&&\frac{\muhad^2}{32\pi^2
F_{0}^2}\
\frac{(4\pi\mu^2/\muhad^2)^{\epsilon/2}}{\Gamma(2-\epsilon/2)}\!\!
\int_{0}^{\infty}\!\!\!\!\!\!\!dz
\ z^{-\epsilon/2} W(z)\,,
\eea }
where $\muhad^2$ is a scale used for normalization, e.g. $z\equiv
Q^2/\muhad^2$, but otherwise totally arbitrary. Therefore the apparent 
quadratic dependence of Eq. (\ref{GDELTAS2}) on $\muhad$ is actually 
fictitious. 
There will be nevertheless 
a certain $\muhad$ dependence in the final answer for $\hat B_K$
due to the fact that there is always a residual $\mu$ dependence because 
the effective Lagrangian of Eq. (\ref{effhal}) is constructed only 
to next-to-leading order in $\als$
(see Eqs.(\ref{wilson},\ref{wilsonN})) and we will identify $\mu=\muhad$ as a
typical hadronic scale. For the final numerical applications we shall take 
$1\,\GeV\le\muhad\le 3\, \GeV$.

The divergent (i.e. ill-defined) integral defining 
$g_{\Delta S=2}(\mu,\epsilon)$ in Eq. (\ref{GDELTAS2}) {\it must} be understood
within the same regularization procedure and conventions used to compute
$C_{\Delta S=2}(\mu)$ in Eq. (\ref{wilsonN}), i.e. ${\overline {MS}}$ and the
definition of evanescent operators of ref. \cite{BW90}.

The function $W(z)$ is defined as 
\be\label{doubleviu}
W(z)=z\frac{\muhad^2}{F_0^2} \cW^{(1)}_{LRLR}(z\muhad^2)\ ,
\ee
where $\cW^{(1)}_{LRLR}$ is defined through an integral over the solid angle
of the momentum $q$, 
{\setl \bea\label{intLRLR}
&&\int
d\Omega_{q}\
g_{\mu\nu}\cW_{LRLR}^{\mu\alpha\nu\beta}(q,l)\vert_{\mbox{\rm
{\scriptsize unfactorized}}}=\nn \\
&&\qquad \left(\frac{l^{\alpha}l^{\beta}}{ l^2}-g^{\alpha\beta}\right)\
\cW^{(1)}_{LRLR}(Q^2)\, ,
\eea}
and $Q^2\equiv -q^2$. In Eq. (\ref{intLRLR}) the function 
$\cW_{LRLR}^{\mu\alpha\nu\beta}(q,l)$ stands for 
{\setl \bea\label{LRLR}
&&\cW_{LRLR}^{\mu\alpha\nu\beta}(q,l)=i^3 \lim_{l\ra 0}   
\! \int \!\!d^4x d^4y d^4z
\ e^{i(q\cdot x + l\cdot y - l\cdot z)}\nn \\
&&\langle 0\vert T\{
L_{\bar{s}d}^{\mu}(x)\ R_{\bar{d}s}^{\alpha}(y)
\ L_{\bar{s}d}^{\nu}(0)\ R_{\bar{d}s}^{\beta}(z)\}\vert 0\rangle\quad ,
\eea }
with 
\be
L_{\bar{s}d}^{\mu}(x)=\bar{s}(x)\gamma^{\mu}\gL d(x) \ ,
\ee
and
\be
R_{\bar{d}s}^{\mu}(x)=\bar{d}(x)\gamma^{\mu}\gR s(x)\ .
\ee

Since $W(z)\sim \cO(N_c^0)$ and $F_0^2\sim \cO(N_c)$ one sees that 
Eq. (\ref{GDELTAS2}) yields $g_{\Delta S=2}(\mu,\epsilon) = 1- \cO(1/N_c)$
with the subleading term governed by $W(z)$.

What do we know about $W(z)$ ?  In general $W(z)$ is a complicated 
function of the variable $z$. However in the large-$N_c$ limit it simplifies
somewhat and can be expressed as an infinite sum of single, double and triple 
poles\footnote{J. Bijnens and J. Prades are thanked for pointing out
  the existence of triple poles, which were actually ignored in the previous
  version of this analysis. See Ref. \cite{BK}.} in the form   
{\setl \bea
\label{largeNansatz}
&&W[z]=6-\sum_{i=1}^{\infty}\frac{\alpha_{i}}{\rho_{i}}-\sum_{i=1}^{\infty}
\frac{\beta_{i}}
{\rho_{i}^2}\nn \\
&& +\sum_{i=1}^{\infty}\left\{\frac{\alpha_{i}}{(z+\rho_{i})}+
\frac{\beta_{i}}{(z+\rho_{i})^2} + 
\frac{\gamma_i}{(z+\rho_{i})^3}\right\} ,
\eea }
where $\rho_{i}=M_{i}^2/\muhad^2$ and $M_{i}$ is the mass of the $i$--th
zero-width resonance. If one knew all the masses and constants
appearing in $W(z)$ from the solution to large-$N_c$ QCD one could use this
information, plug it in Eq. (\ref{GDELTAS2}) and compute $\hat B_K$ through 
Eq. (\ref{BKhat}). Since this solution is not available, we will have to 
follow a different strategy. Firstly, one recalls 
that it is due to the presence of a perturbative contribution 
to a given Green's function 
that one must take these sums over resonances truly up to
infinity because otherwise parton-model logarithms cannot be
reproduced\cite{Wi79}. 
Clearly, when the sums extend all the way to infinity, in
general one cannot expand in powers of $1/z$ at large $z$ in a naive way.
However our function $W(z)$ is an order
parameter of spontaneous chiral symmetry breaking or, in other words, the
contribution of perturbation theory to $W(z)$ vanishes {\it to all orders}. 
One can then try to match a naive expansion in $1/z$ to the Operator 
Product Expansion considering  finite sums as an 
approximation to the originally infinite sums. This way one can 
fix the unknown parameters and masses through the knowledge on the
behavior of $W(z)$ at large $z$ given by this Operator Product Expansion. 

We have no proof that these infinite sums have to behave as if they were
finite but we know that, if this is so, a successful
framework such as Vector Meson Dominance then becomes the leading term in the  
approximation to large $N_c$ that we propose, instead of 
a phenomenological ansatz\cite{VMD1}. Besides, our approximation works
well in those cases where we can compare with the experiment\cite{VMD0,VMD2}. 

Consequently we impose the following behavior on $W(z)$: 
\begin{itemize}
\item
i)At low $z$, from chiral perturbation theory, one finds 
\cite{GF95,HKS99,BP99,BK}  
{\setl \bea \label{Wzlowenergy}
&&W(z)\approx \quad 6\nn \\
&&- 24 \frac{\muhad^2}{F_0^2}\left(2 L_1+5L_2+L_3+L_9\right)\ z \\ 
&&\qquad \qquad + \dots \nn
\eea}
\item
ii)At high $z$, from the operator product expansion, one gets\cite{BK}
{\setl 
\bea\label{Wzhighenergy}
&&W(z)\approx \nn \\
&&\frac{24 \pi \als F_0^2}{\muhad^2  } 
\left[1\!\! +\! \frac{\epsilon}{12}\left(5\!+ \kappa\right)\!+
  \cO(\als) \right]\frac{1}{z} \nn \\
&&\qquad \qquad +\dots 
\eea}
\end{itemize}
Notice that in i) chiral logarithms are neglected, in accord with the
large-$N_c$ expansion. It is a phenomenological fact that they are small in
vector and axial-vector channels, which are precisely the relevant ones here.
For the $L_i$ couplings we use lowest meson dominance, as explained in 
ref. \cite{VMD0,VMD1}. This also works
phenomenologically. The factor of 6 in Eq. (\ref{Wzlowenergy}) explains the
corresponding factor in $W(z)$ in Eq. (\ref{largeNansatz}). 

As to ii), notice the $\epsilon$ and $\kappa$ dependence of the result. This
is brought about by the need to do the calculation in d dimensions, as
repeatedly emphasized. Then the appearance of structures like
$\gamma_{\lambda} (1-\gamma_5)\gamma_{\rho}\gamma_{\mu}\otimes 
\gamma^{\lambda} (1-\gamma_5)\gamma^{\rho}\gamma^{\mu}$ in the OPE calculation
of $W(z)$ makes the definition of
$\gamma_5$ and evanescent operators of practical relevance. We followed
closely refs. \cite{BW90,BBL96}.

Considering the higher orders in Eq. (\ref{Wzhighenergy}) 
there has
recently been the warning, issued by the authors of ref. \cite{warning}, that
these subleading effects in the OPE expansion might actually be numerically
important. I don't expect this to be the case here because, as we shall see,
the impact of the OPE fall-off in the final numerical value is not large, 
but a more detailed analysis is currently underway\cite{BKnew}.

All in all, Eqs.(\ref{Wzlowenergy},\ref{Wzhighenergy}) 
impose the following conditions on the parameters and
masses of $W(z)$:
\bea\label{slope}
&&\sum_{i=1}^{\infty}\frac{\alpha_{i}}{\rho_{i}^2}+2\sum_{i=1}^{\infty}
\frac{\beta_{i}}{\rho_{i}^3} + 3 \sum_{i=1}^{\infty}
\frac{\gamma_{i}}{\rho_{i}^4} =\nn\\
&&24\frac{\muhad^2}
{F_{0}^2}[2L_{1}+5L_{2}+L_{3}+L_{9}]\,
\eea
from $\cO(z)$ terms in Eq. (\ref{Wzlowenergy}). Furthermore,  
\be\label{Q2}
\sum_{i=1}^{\infty} 
\frac{\alpha_{i}}{\rho_{i}}+\sum_{i=1}^{\infty}\frac{\beta_{i}}
{\rho_{i}^2} + \sum_{i=1}^{\infty} \frac{\gamma_{i}}{\rho_{i}^3} = 6\, 
\ee
from the absence of $\cO(1/z^0)$ terms in Eq. (\ref{Wzhighenergy}); also    
\be
\sum_{i=1}^{\infty}\alpha_{i}=\cR+\frac{\epsilon}{2}\cS\,, 
\ee
with
{\setl \bea
&&\cR=\left[24\pi^2\frac{\als}{\pi}+\cO\left(\frac{N_c
\als^{2}}{\pi^{2}}\right)\right]
\frac{ F_{0}^2}{\muhad^2}\quad \nn \\
&&\cS=\left[4\pi^2
(5+\kappa)\frac{\als}
{\pi}+\cO\left(\frac{N_c\als^{2}}{\pi^{2}}\right)\right]
\frac{F_{0}^2}{\muhad^2}\quad , 
\eea}
from the $\cO(1/z)$ term in Eq. (\ref{Wzhighenergy}).

\begin{figure*}
\centering
\epsfig{file=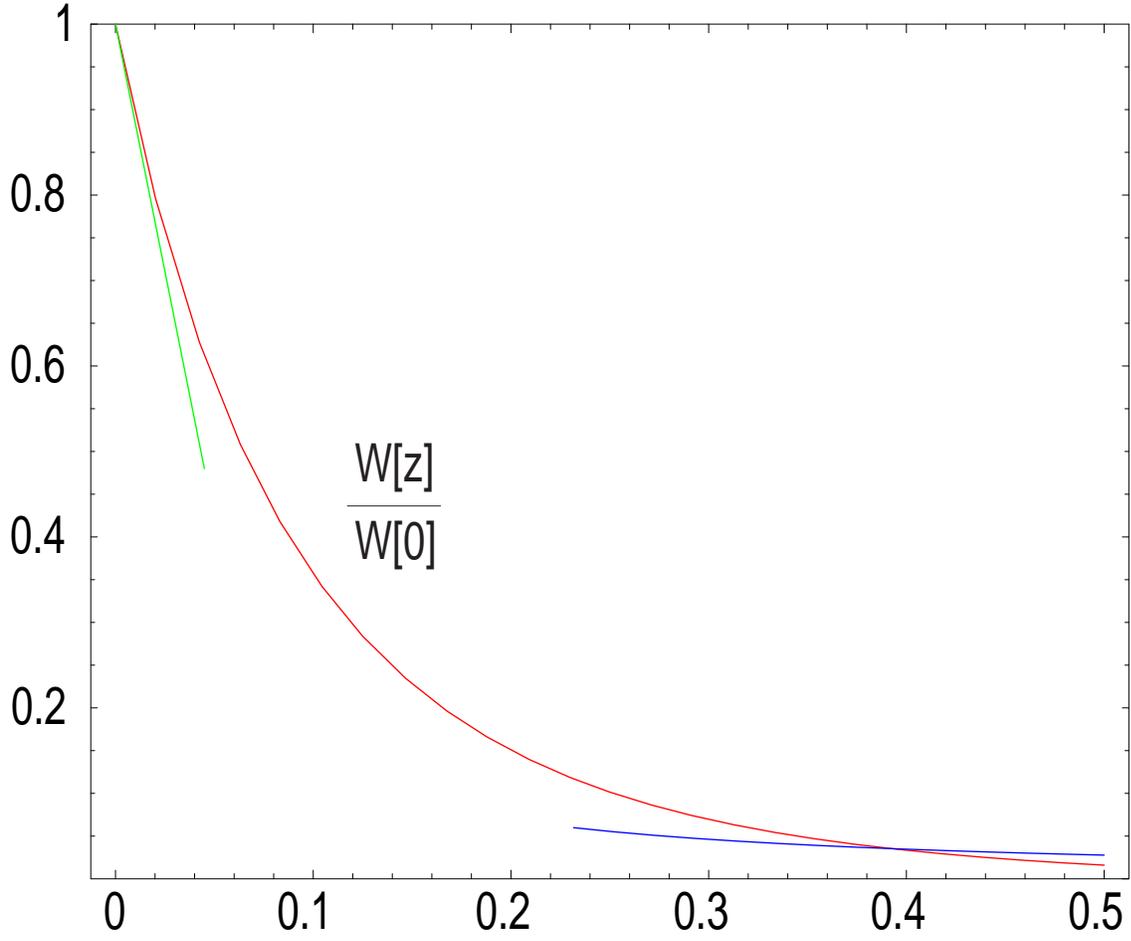}

\caption{\it Plot of the hadronic function $W[z]$ in 
Eq.~(\ref{largeNansatz}) versus $z=Q^2/\muhad^2$ for $\muhad=1.4$ GeV,
$F_0=85.3$ MeV and $M_V= 770$ MeV. 
The red curve is the function
$W[z]$ corresponding to the contribution of a single $\rho$ resonance 
(see text). The green line represents the low--energy chiral 
behaviour and the blue curve the
prediction from the OPE.}
\end{figure*}

When all these constraints are imposed on $W(z)$ in 
Eq. (\ref{largeNansatz}) the expression for $g_{\Delta
  S=2}(\mu)$ in Eq. (\ref{GDELTAS2}) in the ${\overline {MS}}$ 
renormalization scheme reads  
{\setl \bea
&&g_{\Delta
S=2}^{(r)}(\mu)  =  1-\frac{\muhad^2}{32\pi^2
F_{0}^2}\Bigg\{\cR
\log \frac{\mu^2}{\muhad^2}+\!\!\cR\!+\!\!\cS \nn \\
&&\quad \quad +\sum_{i=1}^{\infty}\left[-\alpha_{i}\log\rho_{i}+
\frac{\beta_{i} }{\rho_{i}}+ \frac{1}{2}\frac{\gamma_{i}}{\rho_{i}^2} \right] 
\Bigg\} \ ,\nn \\
& \Ra & \left(\frac{\als(\mu)}{\als(\muhad)}\right)^{\frac{9}{11}\frac{1}{N_c}}
\!\!\Bigg\{1-\frac{1}{2}\frac{\muhad^2}{16\pi^2
F_{0}^2}\Bigg(\!\cR \!+\!\cS\! \nn \\
&&\quad \quad +\sum_{i=1}^{\infty}\left[-\alpha_{i}\log\rho_{i}+
\frac{\beta_{i} }{\rho_{i}}+ \frac{1}{2}\frac{\gamma_{i}}{\rho_{i}^2}
\right]\Bigg) \Bigg\} \,,
\eea}
where in the second line we have written the renormalization group-improved
result. We would like to insist on the fact 
that this result, contrary to the various
large--$N_c$  inspired calculations which have been published so
far~\cite{BBG88,G90,PdeR91,GF95,BEFL98,HKS99,BP99} is the  
{\it full} result  to next--to--leading order in the $1/N_c$
expansion. The renormalization $\mu$ scale dependence as well as the
scheme dependence in the factor
$\cS$ cancel exactly, at the next--to--leading log approximation, 
with the short distance $\mu$ and scheme dependences in the Wilson
coefficient
$C_{\Delta S=2}(\mu)$ in Eq.~(\ref{wilsonN}). Therefore, the {\it full}
expression of the invariant
$\hat{B}_{K}$--factor to lowest order in the chiral expansion and to
next--to--leading order in the $1/N_c$ expansion, which takes 
into account the hadronic contribution from light quarks below a mass scale
$\muhad$ is then  
{\setl\bea\label{finalBK}
&&\hat{B}_{K} = 
\left(\frac{1}{\als(\muhad)}\right)^{\frac{3}{11}}\frac{3}{4}\nn \\
&&\left\{1  - 
\frac{\als(\muhad)}{\pi}\frac{1229}{1936}+\cO\left(\frac{N_c
\als^{2}}{\pi^{2}}\right)\right. \nn \\
 &  &-\left.
\frac{\muhad^2}
{32\pi^2
F_{0}^2}
\sum_{i=1}^{\infty}\left[-\alpha_{i}\log\rho_{i}+
\frac{\beta_{i} }{\rho_{i}} + \frac{1}{2}\frac{\gamma_{i} }{\rho_{i}^2}\right]
\right\}\,.
\eea}
As remarked earlier, 
the numerical choice of the hadronic scale $\muhad$ is, a priori, arbitrary.
In practice, however, $\muhad$ has to be sufficiently large so as to make
meaningful the truncated pQCD series in the first line. Our final error in the
numerical evaluation of
$\hat{B}_{K}$ will include the small effect of fixing $\muhad$ within the
range $1\, \GeV\le\muhad\le 3\, \GeV$.   

How many resonances should one include in the sum in Eq. (\ref{finalBK})?
Clearly, the more the better. However, 
given that there is only a limited amount of information coming 
from Eqs. (\ref{Wzlowenergy}, \ref{Wzhighenergy})
one can in practice determine the couplings of just one resonance, the $\rho$
vector meson. With them, and given the rho mass, the
  corresponding $W(z)$ function of Eq. (\ref{largeNansatz}) is shown in
  Fig. 1, where it is seen that $W(z)$ intersects the OPE
  curve at a certain value of $z$ that we will call $\hat z$. It turns out
  that  $\hat z\sim 0.39$ (i.e. $Q\sim 900 $ MeV). As can be seen, the
matching to the OPE curve is not perfect and this is why, for the final
numerical estimate, we propose to use the hadronic curve $W(z)$ only in
the region $0\leq z\leq \hat z$ while in the range $\hat z\leq z < \infty$ we
use the first term of the OPE, Eq. (\ref{Wzhighenergy}). 
This results in a different expression for the 
$\hat B_K$ parameter which reads
{\setl
\bea\label{ansatzBK}
&&\hat{B}_{K}\left(\rho,a_1,\rho'\right)   =  
\left(\frac{1}{\als(\muhad)}\right)^{\frac{3}{11}}\frac{3}{4}\nn \\
&&\Bigg\{\!1\!\!  -\! 
\frac{\als(\muhad)}{\pi}\left[\frac{1229}{1936}-\!
\frac{3}{4}\!\log\hat{z}\right]\!+\!
\cO\left(\frac{N_c \als^{2}}{\pi^{2}}\right)\nn \\
&&
-\frac{\muhad^2}
{32\pi^2
F_{0}^2}\!\!\Big[\alpha_{V}
\log\frac{\hat{z}+\rho_{V}}{\rho_{V}} 
- \beta_{V}\left(\frac{1}{\hat{z}+\rho_{V}}-
\frac{1}{\rho_{V}}\right)\nn \\ 
&& - \frac{\gamma_V}{2} 
\left(\frac{1}{(\hat{z}+\rho_{V})^2}-
\frac{1}{\rho_{V}^2}\right) \Big]\Bigg\}\,.
\eea}
Of course in the limit $\hat z \ra \infty$, Eq. (\ref{ansatzBK}) reproduces 
Eq.(\ref{finalBK}), but this limit in practice can only be achieved by using
more and more resonances in the sum, which requires the knowledge of more and
more terms in both the low and high-energy expansions of $W(z)$.
The physical interpretation of the result (\ref{ansatzBK}) 
is clear: the area under the curve in Fig. 1 is
roughly the amount one subtracts from the large-$N_c$ limit result of $\hat
B_K =3/4$. In numbers one gets
\be\label{finalnumber}
\hat{B}_{K}=0.38\pm 0.11 \,,
\ee
where the error is the combined limit of errors from short-- (e.g. how much is
$\Lambda_{\overline{MS}}$?) and
long--distance (e.g. how much is the $\rho$ mass?)
 contributions to $\hat{B}_{K}$ for a choice of $\muhad$ in the
range
$1\,\GeV\le\muhad\le 3\,\GeV$. I refer the reader to ref. \cite{BK} for
details.  
This result, of course, does not include the error due to next--to--next--to
leading terms in the
$1/N_c$ expansion, nor the error due to chiral corrections in the unfactorized
contribution, but we consider it to be a rather
{\it robust} prediction of $\hat{B}_{K}$ in the chiral limit and at the
next--to-- leading order in the $1/N_c$ expansion for the following reasons. 
Firstly, Fig. 1 shows that most of the area under the curve is in the region
$0\leq z\lesssim 0.4$ which makes the final number for $\hat B_K$ relatively
insensitive to the part of the curve coming from the OPE. Secondly, at a more
qualitative level, if one 
accepts 
that the low-$z$ region is correctly given by Chiral Perturbation Theory up to
a scale of $z\sim 0.06$ (i.e. $E\sim 350$ MeV) and the OPE also describes
reasonably well the high-$z$ region starting from 
$z\sim 0.5$ ($E\sim 1$ GeV); then the conclusion is that there isn't much room
left in between to do a lot of things !  

Our final result in Eq.(~\ref{finalnumber}) is compatible with 
the current algebra prediction~\cite{DGH82} obtained from 
the $K^{+}\ra \pi^{+}\pi^{0}$ decay rate at lowest order in chiral
perturbation theory. 
It turns out~\cite{PdeR96} that the bosonization of the four--quark
operator $Q_{\Delta S=2}$ and the bosonization of the operator $Q_{2}-Q_{1}$
which generates $\Delta I=1/2$ transitions  are related to each other in
the combined chiral limit and  next--to--leading order $1/N_c$ expansion.
It follows that decreasing the value of $\hat{B}_{K}$ from the 
large--$N_c$ prediction of 3/4 down to the result in 
Eq.~(\ref{finalnumber}) is correlated with
an increase of the coupling constant in the lowest order Effective Chiral
Lagrangian which generates $\Delta I=1/2$ transitions, providing therefore a
first glimpse at a quantitative understanding of the dynamical origin of
the  $\Delta I=1/2$ rule. In the future we would like 
to make some refinements and 
a detailed comparison between our result and other previous
estimates\cite{BKnew}. Among these ``refinements''
one has most notably the inclusion of higher order terms in the chiral
expansion. These are crucial in order to be able to make a meaningful
comparison with the values favoured by lattice QCD
determinations~\cite{Ku00} as well as by recent phenomenological
analysis~\cite{Cietal99,CPRS00}.

I would not like to finish without stressing that our framework {\it can} in
principle be improved systematically, for instance by computing more terms in
Eqs. (\ref{Wzlowenergy},\ref{Wzhighenergy}). It can be applied as well 
to other physical situations involving weak matrix
elements. In order to do this one might find helpful the following

\vskip 0.5 cm

{\bf DO-IT-YOURSELF KIT:}

\begin{itemize}
\item
i)Pick the QCD Green's function, $\cG$, 
which defines the matrix element you are
interested in.
\item
ii)Find the large-$N_c$ resonance 
representation of $\cG$ (i.e. the interpolating function).
\item
iii)Impose on the above interpolating function 
the long- and short-distance constraints stemming from
Chiral Perturbation Theory and the Operator Product Expansion, respectively.
\item
iv)Compute number.
\item
v)Enjoy.
\end{itemize}
         
\vspace*{7mm} 
{\large{\bf Acknowledgments}}

\vspace*{3mm}

\noi
I thank E. de Rafael for a most enjoyable collaboration. 
During the elaboration of this work we benefited a great deal from 
discussions with Marc Knecht. We also thank 
Hans Bijnens, Maarten Golterman, Michel Perrottet, Toni Pich and 
Ximo Prades for discussions. (S.P.) is 
also grateful to C. Bernard and Y. Kuramashi
for conversations  and to the Physics Dept. of Washington University
in  Saint Louis for the hospitality extended to him while the work of
ref. \cite{BK} was being finished.
This work has been supported in part by
TMR, EC-Contract No. ERBFMRX-CT980169 (EURODA$\phi$NE) and by the 
research project CICYT-AEN99-0766.
I thank S. Narison for his kind invitation to the Conference.



{\bf DISCUSSION}

{\bf H. Fritzsch (Munich)}

I would expect that you encountered problems with your technique if the singlet
$0^{-+}$ is involved since the role of the gluon anomaly is different for
large $N_c$.

{\bf S.P.}

As is well known, 
the gluon anomaly is at the origin of the $\eta'$ mass. In the strict 
large $N_c$ limit the anomaly is switched off and $M_{\eta'}\ra 0$. It may
seem then that the large-$N_c$ expansion is a very bad expansion 
for the $\eta'$ since in the real world 
$ M_{\eta'}\sim M_{\rm proton}\sim 1$ GeV $>> M_{\pi}$; i.e. it doesn't look 
at all like a Goldstone boson which is what the $N_c=\infty$ result says.

Although I don't know whether the $1/N_c$ expansion will be in the end 
really helpful for 
calculations involving the $0^{-+}$ channel, I would like to explain why I
think that the above argument is potentially fallacious. Take 
good old QED as an expansion in powers of $\alpha$ around $\alpha=0$. If you
were to compare the real world (where e$^-$ and e$^+$ certainly scatter) 
to what you
find from the strictly speaking first term in this expansion 
, which is the one at $\alpha=0$ (i.e. no scattering), you
could also conclude that the $\alpha$ expansion is a disaster. Well, in some
respects $\alpha=0$ corresponds to $N_c=\infty$. In other words, one should
always go to {\it first nontrivial order} in the expansion before comparing to
the real world. It is not unconceivable then
that, after the bulk of the $\eta'$ mass is included in the first
nontrivial term in the $1/N_c$ expansion, the rest of the series in $1/N_c$
turns out to be a well-behaved innocent-looking power series.

Having said this, I also want to stress that in our particular case of $\hat
B_K$, at the order we are computing, 
the $0^{-+}$ channel doesn't play any role at all. It's all governed by
vectors and axial-vectors for which we know that resonance saturation (which is
basically  what the large-$N_c$ expansion amounts to in this case) is a good
thing to do. (I thank M. Knecht for reminding me of this.)


\begin{thebibliography}{99}

\bibitem{Jamin}
         A.J. Buras, M. Jamin and P. Weisz, Nucl. Phys. {\bf B347} (1990) 491;
         S. Herrlich and U. Nierste, Nucl. Phys. {\bf B419} (1994) 292,
         ibid. Nucl. Phys. {\bf B476} (1996) 27.


\bibitem{BBL96}
         G.~Buchalla, A.J.~Buras and M.E.~Lautenbacher, 
         Rev. Mod. Phys. {\bf 68} (1996) 1125;
         A.J.~Buras, in Les Houches Lectures, Session LXVIII, 
         {\it Probing the
         Standard Model of Particle Interactions}, eds. R.~Gupta, A.~Morel,
         E.~de Rafael and F.~David, North--Holland 1999. 

\bibitem{PdeR91}
         A.~Pich and E.~de Rafael, Nucl. Phys. {\bf B358} (1991) 311.  

\bibitem{deR95}
         E.~de Rafael, ``Chiral Lagrangians and Kaon CP--Violation'', in
         {\it CP Violation and the Limits of the Standard Model}, Proc.
         TASI'94, ed. J.F.~Donoghue (World Scientific, Singapore, 1995)

\bibitem{tH74}
         G~'t Hooft, Nucl. Phys. {\bf B72} (1974) 461; {\bf B75} (1974) 461.

\bibitem{Wi79}
         E.~Witten, Nucl. Phys. {\bf B79} (1979) 57.

\bibitem{BK}
         S. Peris and E. de Rafael, Phys. Lett. {\bf B490} (2000) 213. A
         missing term in the function $W[z]$ of this reference has been
         corrected in hep-ph/0006146 v3.

\bibitem{BW90}
         A.J. Buras and P.H. Weisz, Nucl. Phys. {\bf B333} (1990) 66.

\bibitem{VMD1}
         M. Knecht, S. Peris and E. de Rafael, Nucl. Phys. (Proc. Suppl.) {\bf
         B86} (2000) 279; see also 
         M. Knecht and E. de Rafael, Phys. Lett. {\bf B424} (1998) 335;
         S. Peris, M. Perrottet and E. de Rafael, JHEP {\bf 05} (1998) 011;
         M. Golterman and S. Peris, Phys. Rev.{\bf D61} (2000) 034018.

\bibitem{VMD0}
         G.~Ecker, J.~Gasser, A.~Pich and E.~de Rafael, Nucl. Phys.  
         {\bf B321} (1989) 311; 
         G.~Ecker, J.~Gasser, H.~Leutwyler, A.~Pich and E.~de Rafael, Phys.
         Lett. {\bf B223} (1989) 425.

\bibitem{VMD2}
         S. Peris, B. Phily and E. de Rafael, hep-ph/0007338; 
         M. Knecht, S. Peris, M. Perrottet and E. de Rafael, Phys. Rev. 
         Lett. {\bf 83} (1999) 5230; see also M. Kecht, S. Peris and E. 
         de Rafael, Phys. Lett. {\bf B443} (1998) 255.

\bibitem{GF95}
         J.P.~Fatelo and J.-M.~G\`{e}rard, Phys. Lett. 
         {\bf B347} (1995) 136.

\bibitem{HKS99}
         T.~Hambye, G.O.~K\"{o}hler and P.H.~Soldan, Eur. Phys. J. 
         {\bf C10} (1999) 271.

\bibitem{BP99}
         J.~Bijnens and J.~Prades, Phys. Lett. JHEP {\bf 01} (1999) 023.

\bibitem{warning}
         V. Cirigliano, J.F. Donoghue and E. Golowich, hep-ph/0007196;  
         J.F. Donoghue, these proceedings.

\bibitem{BKnew}
         S. Peris and E. de Rafael, in preparation.

\bibitem{BBG88}
         W.A.~Bardeen, A.J.~Buras and J.-M.~G\`erard, Phys. Lett. 
         {\bf B211} (1988) 343.

\bibitem{G90}
         J.-M.~G\`{e}rard, Acta Physica Polonica {\bf B21} (1990) 257.

\bibitem{BEFL98}
         S.~Bertolini, J.O.~Egg, M.~Fabbrichesi and E.I.~Lashin, Nucl. Phys. 
         {\bf B514} (1998) 63. 

\bibitem{DGH82}
         J.F.~Donoghue, E.~Golowich and B.R.~Holstein, Phys. Lett. 
         {\bf B119} (1982) 412.

\bibitem{PdeR96}
         A.~Pich and E.~de Rafael, Phys. Lett. {\bf B374} (1996) 186.

\bibitem{Ku00}
         Y.~Kuramashi, Nucl. Phys. B (Proc. Suppl.) 83-84 (2000) 24.         

\bibitem{Cietal99}
         M. Ciuchini, E. Franco, L. Giusti, V. Lubicz, G. Martinelli, 
         Nucl. Phys. {\bf B573} (2000) 201. 

\bibitem{CPRS00}
         F.~Caravaglios, F.~Parodi, P.~Roudeau and A.~Stocchi, hep-ph/0002171. 

\end{thebibliography}
\end{document}